\documentclass[fleqn]{annalen}
\usepackage{latexsym,amssymb}
\usepackage{epsf}
\usepackage{supercite}
\begin{document}
\newcommand{\volume}{8}              
\newcommand{\xyear}{1999}            
\newcommand{\issue}{5}               
\newcommand{\recdate}{29 July 1999}  
\newcommand{\revdate}{dd.mm.yyyy}    
\newcommand{\revnum}{0}              
\newcommand{\accdate}{dd.mm.yyyy}    
\newcommand{\coeditor}{ue}           
\newcommand{\firstpage}{1}         
\newcommand{\lastpage}{10}          
\setcounter{page}{\firstpage}        
\newcommand{\keywords}{two interacting particles, disordered solids, metal-insulator transitions}
\newcommand{\PACS}{61.43.--j, 71.30.+h, 72.15.Rn}
\newcommand{\shorttitle}{R. A. R\"{o}mer et al., Two interacting particles in a random environment} 
\title{Numerical results for two interacting particles\\in a random environment}
\author{R.\ A.\ R\"{o}mer$^{1}$, M.\ Leadbeater$^{2}$, and M.\ Schreiber$^{1}$}
\newcommand{\address}
  {$^{1}$Institut f\"{u}r Physik,
  Technische Universit\"{a}t, D-09107 Chemnitz, Germany, \\
  $^{2}$Dipartimento di Fisica,
  Univ.\ di Roma Tre, Via delle Vasca Navale 84, I-00146 Roma, Italy
}
\newcommand{\email}{\tt r.roemer@physik.tu-chemnitz.de}
\maketitle
\begin{abstract}
  Much evidence has been collected to date which shows that repulsive
  electron-electron interaction can lead to the formation of particle
  pairs in a one-dimensional random energy landscape. The localization
  length $\lambda_2$ of these pair states is finite, but larger than
  the localization length $\lambda_1$ of the individual particles.
  After a short review of previous work, we present numerical evidence
  for this effect based on an analysis of the interaction matrix
  elements and an application of the decimation method. The results
  based on the decimation method for a two-dimensional disordered
  medium support the localization-delocalization transition of pair
  states predicted recently.
\end{abstract}
\newcommand{\figwidth}{0.68\textwidth}
\newcommand{\fullfigwidth}{1.2\textwidth}
\newcommand{\halffigwidth}{0.4\textwidth}
\newcommand{\halffigwidthb}{0.43\textwidth}

\section{Introduction}
\label{sec-intro}

Investigations into the interplay of disorder and many-body
interactions continue to receive a lot of attention mainly due to the
persistent current problem \cite{ChaWBK91,RomP95,SchSSS98} and the
experimental discovery of the two-dimensional (2D)
metal-insulator-transition (MIT) \cite{KraKFP94,KraMBF95,KraSSM96}.
In order to theoretically study this interplay, in principle with
increasing system size one has to solve a problem with an
exponentially growing number of states in the Hilbert space. At
present, this can be achieved only for a few particles in 1D
\cite{Whi93,SchJWP98} and very few particles in 2D
\cite{VojES98a,VojES98b,She99,SonS99}. Shepelyansky
\cite{She94,She96a} --- following earlier work of Dhorokov
\cite{Dor90} --- proposed another approach looking at the properties
of {\em two} interacting particles (TIP) in a random environment. The
TIP Hamiltonian is
\begin{eqnarray}
{\bf H} & = &
- t \sum_{n,m} \left( \vert n, m \rangle\langle n+1,m\vert +
                      \vert n, m \rangle\langle n,m+1\vert + h.c. \right)
\nonumber \\
 & & \mbox{ }
+ \sum_{n, m} \vert n, m \rangle
              \left( \epsilon_n + \epsilon_m + U\delta_{nm} \right)
              \langle n,m\vert
\label{eq-ham}
\end{eqnarray}
with positions $n$, $m$ of each particle on a chain of length $L$,
hopping probability $t$ (which we use to define the energy scale, i.e., $t=1$), onsite
interaction strength $U$, and random potentials $\epsilon_n \in
[-W/2,W/2]$ for disorder $W$.  Shepelyansky suggested that even for
repulsive interactions the two particles would form pairs with larger
localization length $\lambda_2$ than the two separate single particles
(SP).  Thus the interaction would enhance the possibility of transport
through the system \cite{Imr96}.  The perhaps even more surprising
prediction was that at pair energy $E=0$
\begin{equation}
  \lambda_2 \propto U^2
  {\lambda_1}^2,
\label{eq-lambda2}
\end{equation}
where $\lambda_1$ is the SP localization length. Since $\lambda_1
\propto 105/W^2$ in 1D, this implies large values of $\lambda_2$ for
small disorders $W$.

The first numerical studies devoted to the TIP problem used the
transfer matrix method (TMM) to investigate the proposed enhancement
of $\lambda_2$
\cite{She94,FraMPW95,RomS97a,FraMPW97,RomS97b,RomS98,Hal96T,HalMK98,HalZMK98}.
Other direct numerical approaches to the TIP problem have been based
on the time evolution of wave packets
\cite{She94,BorS95,BorS97,BriGKT98,HalKK98}, exact diagonalization
\cite{WeiMPF95}, variants of energy-level statistics
\cite{WeiP96,AkkP97,HalZMK98b} and multifractal analysis
\cite{WaiWP99,WaiP98}, Green function approaches
\cite{OppWM96,SonK97,SonO98,Fra98}, perturbative methods
\cite{JacS95,JacSS97} and mappings to effective models
\cite{Imr95,FraM95,FraMP96,FraMP97,PonS97}.  In these investigations
usually (but not always
\cite{FraMPW95,RomS97a,FraMPW97,RomS97b,RomS98,Hal96T,HalMK98,HalZMK98})
an enhancement of $\lambda_2$ compared to $\lambda_1$ has been found
but the quantitative results tend to differ both from the analytical
prediction in Eq.\ (\ref{eq-lambda2}), and from each other.
Furthermore, a check of the functional dependence of $\lambda_2$ on
$\lambda_1$ is numerically very expensive since it requires very large
system sizes $L \gg \lambda_2 \gg \lambda_1$.  Extensions of the
original arguments have been proposed for higher dimensions
\cite{Imr96,Imr95,BorS97}, TIP close to a Fermi sea \cite{OppW95} and
long-range interactions in 1D \cite{RomS97a,RomS98,BriGKT98} and 2D
\cite{She99}.

In this paper, we present numerical results for the interaction matrix
elements and show that a correct statistical description necessitates
the use of the {\em logarithmic} instead of the commonly used
arithmetic average. The resulting {\em typical} matrix element $u_{\rm
  typ}$ has a different disorder dependence than the arithmetic mean
$u_{\rm abs}$.  Following the arguments of Ref.\ \cite{She94,She96a},
the use of $u_{\rm typ}$ does not give rise to a power-law enhancement
as in Eq.~(\ref{eq-lambda2}). However, taking into account the energy
denominators, the power-law can be recovered but with a smaller power
of, e.g., $1.4\pm 0.2$ for $U=1$. This value is in agreement with
previous results in 1D
\cite{FraMPW95,RomS98,Hal96T,HalMK98,HalZMK98,BriGKT98,HalKK98,WeiP96,AkkP97,WaiWP99,WaiP98,OppWM96,SonK97,SonO98,Fra98,JacS95,JacSS97,Imr95,PonS97,LeaRS99}.
We further review results based on the application of the decimation
method \cite{LeaRS99,LeaRS98,RomLS99} which allows for a direct
computation of $\lambda_2(U)$.  Augmenting the analysis with the
finite-size-scaling (FSS) approach \cite{MacK83}, we construct
estimates $\xi_2(U)$ of the TIP localization length in the infinite
system \cite{LeaRS99,RomLS99}. For $U=0$ they reproduce accurately the
well-known dependence of $\lambda_1$ on disorder while for finite $U$,
we find $ \xi_{2}(U) \sim \xi_2(0)^{\beta(U)} $ with $\beta(U)$
varying between $\beta(0)=1$ and $\beta(1) \approx 1.5$.  Thus the
enhancement persists, unlike for TMM
\cite{FraMPW95,RomS97a,FraMPW97,RomS97b,RomS98,HalMK98,HalZMK98}, in
the limit of large system size. Finally, we apply the decimation
method in 2D.  Again using FSS we find scaling functions for $0.5 \leq U
\leq 2$ which exhibit two branches corresponding to localized and
extended behavior for TIP in an infinite system, and we estimate the
$U$ dependence of the critical disorder and the critical exponent at
the transition.


Before presenting our data, we briefly review the analytical approach
to TIP.  The prediction of Shepelyansky is based upon looking at the
interaction matrix element of a pair state $\psi_{kl}=\psi_k\psi_l$
with another pair state $\psi_{nm}$ \cite{She94,She96a}. Here $\psi_k$
denotes the SP eigenstate localized with $\lambda_1$ around site $k$
and we can restrict the states by $|k-l|\leq \lambda_1$, $|n-m|\leq
\lambda_1$ and $|k-n|\leq \lambda_1$, $|l-m|\leq \lambda_1$.  The
interaction matrix elements are
\begin{equation}
u  
  = \langle \psi_{kl} \vert U \vert \psi_{nm} \rangle
  = U \sum_{j=1}^{N} \psi_k^\dagger(j)\psi_l^\dagger(j) \psi_n(j) \psi_m(j),
\end{equation}
where we used the Hubbard onsite interaction \cite{RomP95} $U
\sum_{j=1}^N n_{j\downarrow} n_{j\uparrow}$ with the number operator
$n_{j\sigma}$ at site $j$ and spin $\sigma$. Assuming
\cite{She94,Imr95} that the SP state is given as
\begin{equation}
  \psi_k(j)\propto \frac{1}{\sqrt{\lambda_1}} \exp - \left(
  \frac{|j-k|}{\lambda_1} + i \theta_j\right)
\label{eq-psi1}
\end{equation}
with $\theta_j$ a random phase, one finds \cite{She94} that the
average of $u$ has a magnitude of
\begin{equation}
u_{\rm abs} \propto \lambda_1^{-3/2}.
\label{eq-u}
\end{equation}
\begin{figure}[t]
  \epsfxsize=\figwidth \centerline{\epsffile{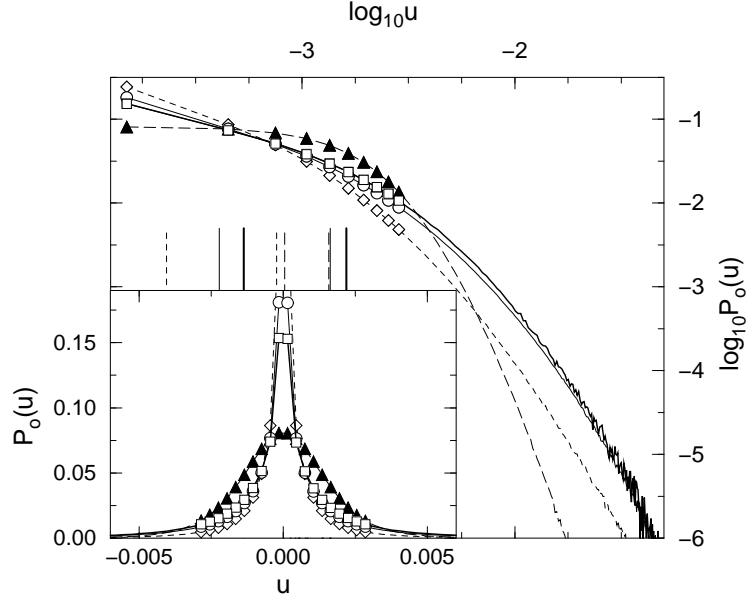}}
 \caption{\label{fig-distribution}
   Normalized distributions $P_{\rm o}(u)$ of the off-diagonal
   coupling matrix elements $u$ for $\lambda_1=26$ ($W=2$), $L=200$,
   and $U=1$. Thin solid, short-dashed, and long-dashed lines
   correspond to the TIP problem, the TIP problem with random
   interaction, and Eq.\ (\protect\ref{eq-psi1}), respectively. The
   thick solid line denote TIP matrix elements for states with energy
   separation less than $\Delta E = 1/2$.  Circles, diamonds,
   triangles, and squares mark the data for the $10$ smallest $|u|$ in
   each case. The vertical lines above the inset indicate the
   respective values of $u_{\rm abs} > u_{\rm typ}$ as computed from
   the total distribution $P(u)$. Inset: same data on a linear scale.}
\end{figure}
Shepelyansky calculated the decay rate $\Gamma$ of a non-interacting
eigenstate by means of Fermi's golden rule $\Gamma \sim U^2/\lambda_1
t$ \cite{She94,She96a,JacS95}. Since the typical hopping distance is
of the order of $\lambda_1$ the diffusion constant is $D \sim U^2
\lambda_1 /t$.  Within a time $\tau$ the particle pair visits $N \sim
U \lambda_1^{3/2} t^{-1/2} \tau^{1/2}$ states.  Diffusion stops when
the level spacing of the visited states is of the order of the
frequency resolution $1/\tau$.  This determines the cut-off time
$\tau^*$ and $\lambda_2 \sim \sqrt{ D \tau^*} \sim (U/t)^2
\lambda_1^2$ in agreement with Eq.\ (\ref{eq-lambda2}). Applicability
of Fermi's golden rule requires $\Gamma \gg t/\lambda_1^2$ which is
equivalent to $U^2 \lambda_1 /t^2 \gg 1$.  This is exactly the
condition for an enhancement of $\lambda_2$ compared to $\lambda_1$.
Alternatively, similar results can be obtained by mapping the model
onto a random-matrix model with entries chosen according to Eq.\ 
(\ref{eq-u}) \cite{She94,FraMP96,FraMP97}.

\section{Numerical results for the interaction matrix elements}
\label{sec-tip-rmm}

The qualitative arguments presented above should be checked for
quantitative accuracy. Even before doing so for Eq.\ 
(\ref{eq-lambda2}), it is already worthwhile to test the validity of
(\ref{eq-u}) and the subsequent arguments.  It was shown
\cite{FraMPW95} that the assumption of a Gaussian distribution of $u$
--- necessary for taking the arithmetic or r.m.s.\ average --- was
oversimplified. We calculated \cite{RomSV99} the dependence of the
averages on $\lambda_1$ and system size.  To this end, we diagonalized
the 1D Anderson model for given $L$ and $W$ and computed $u$ for all
suitable states and many disorder configurations. In
Fig.~\ref{fig-distribution} we show the distribution of $u$ and the
values of $u_{\rm typ}$ and $u_{\rm abs}$.
\begin{figure}[t]
  \epsfxsize=\figwidth \centerline{\epsffile{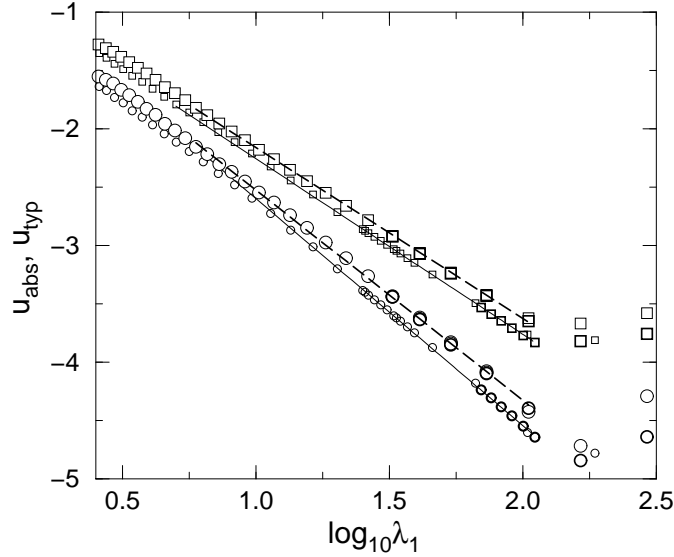}}
 \caption{\label{fig-disdepend}
   Dependence of $u_{{\rm abs}}$ ($\Box$) and $u_{{\rm typ}}$
   ($\circ$) on $\lambda_1$ for the TIP eigenstates from the entire
   band (small symbols) and for $\Delta E= 1/2$ (large symbols) for
   $U=1$ and $L=200$.  The bold symbols indicate $u_{{\rm abs}}$ and
   $u_{{\rm typ}}$ for $L=250$. The solid (broken) lines represent the
   power laws $u_{{\rm abs}} \sim \lambda_1^{-1.5}$
   ($\lambda_1^{-1.45}$) and $u_{{\rm typ}} \sim \lambda_1^{-1.95}$
   ($\lambda_1^{-1.8}$). } 
\end{figure}
Additionally, we include results for an interaction with randomly
varying onsite term $U(j)$ and results for SP states with random phase
as in Eq.\ (\ref{eq-psi1}). Based on these results, it was shown
\cite{RomSV99} that due to the non-Gaussian distribution of $u$, one
should use the logarithmic rather than the arithmetic average for $u$.
However, whereas $u_{\rm abs}\propto \lambda_1^{-1.5}$ \cite{WaiP98},
the logarithmic average obeys $u_{\rm typ}\propto \lambda_1^{-1.95}$
as shown in Fig.~\ref{fig-disdepend}.  Following the arguments above,
this would imply $\lambda_2\propto \lambda_1^{1.1}$.  We emphasize
that this does not mean that there is no enhancement of $\lambda_2$.
Rather, the results \cite{RomSV99} indicate that the arguments of
first-order perturbation theory \cite{She94} capture the physics in a
somewhat oversimplified form. Noting the possibly quite different
energies of the pair states, it appears more suitable to average $u$
only over such states which differ in energy by $\Delta E \leq \pm
|U|/2$.  This leads to a different $\lambda_1$ dependence with reduced
exponent, e.g., $u_{{\rm typ}} \sim \lambda_1^{-1.8}$ for $U=1$, cp.\ 
Fig.~\ref{fig-disdepend}, implying
\begin{equation}
\lambda_2 \propto \lambda_1^{1.4\pm 0.2},
\label{eq-tipme-e-lambda2}
\end{equation}
in good agreement with previous data based on a wide variety of
methods
\cite{FraMPW95,RomS98,Hal96T,HalMK98,HalZMK98,BriGKT98,HalKK98,WeiP96,AkkP97,WaiWP99,WaiP98,OppWM96,SonK97,SonO98,Fra98,JacS95,JacSS97,Imr95,PonS97,LeaRS99}
and also with the results of the next section.

\section{Enhancement of the TIP localization length in 1D}
\label{sec-tip-dm1d}
\begin{figure}[t]
  \epsfxsize=\figwidth \centerline{\epsffile{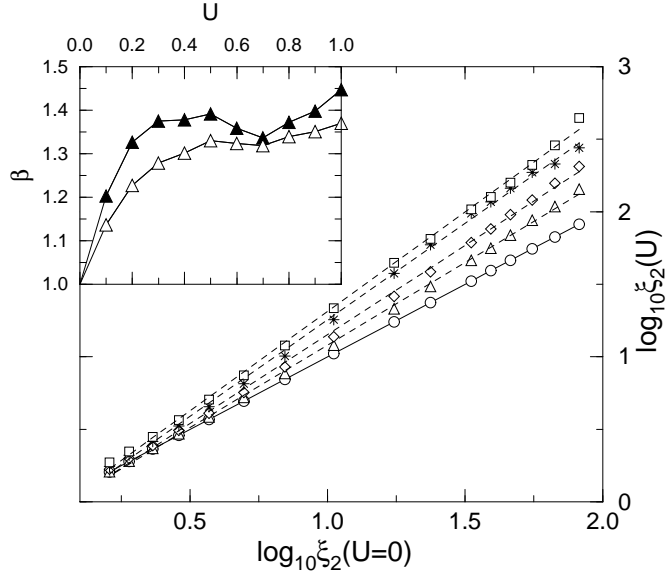}}
 \caption{\label{fig-tipdm-xi0}
   TIP localization lengths $\xi_2$ after FSS for $U=0$ ($\circ$),
   $0.1$ ($\triangle$), $0.2$ ($\Diamond$), $0.5$ ($*$) and $1$
   ($\Box$). The data are for $W \in [1,6]$. The dashed lines show
   fits according to $\xi_2(U) \sim \xi_2(0)^{\beta}$, the solid line
   sets the reference for $U=0$.  Inset: Exponent $\beta$ obtained by
   a power-law fit to the data for $U= 0.1, \ldots, 1$. The filled
   symbols correspond to the fit (\protect\ref{eq-ps}) with finite
   $c$.} 
\end{figure}

The first numerical investigations \cite{She94,FraMPW95} of
$\lambda_2$ used the TMM.  Two of us studied the TIP problem by a
different TMM \cite{RomS97a} at large $L$ and found that (i) the
enhancement $\lambda_2/\lambda_1$ decreases with increasing $L$, (ii)
the behavior of $\lambda_2$ for $U=0$ is equal to $\lambda_1$ in the
limit $L\rightarrow\infty$ only, and (iii) for $U\neq 0$ the
enhancement $\lambda_2/\lambda_1$ also vanishes completely in this
limit.  Therefore we concluded \cite{RomS97a,FraMPW97,RomS97b} that
the TMM applied to the TIP problem in 1D measures an enhancement which
is due to the finiteness of the systems considered.
It is now well understood \cite{SonO98} that the main problem with the
TMM approach is that the enhanced $\lambda_2$ is expected to appear
along the diagonal sites of the TIP Hamiltonian, whereas the
unsymmetrized TMMs \cite{FraMPW95,RomS97a} proceed along an
SP coordinate.  Various new TMM techniques have been developed to take
this into account \cite{She94,FraMPW95,RomS97a,RomS98,HalZMK98}, but
still all TMMs share a common problem: in general the $U=0$ result for
$\lambda_2$ does not equal the value of $\lambda_1/2$ which is
expected for non-interacting particles \cite{LeaRS99}.  Rather,
they yield $\lambda_2(0)\approx\lambda_2( 1 )$ and thus $\lambda_2(0)
\gg \lambda_1/2$. Therefore other methods appear to be more
appropriate for the TIP problem.

Here we briefly review the results \cite{LeaRS99,LeaRS98,RomLS99} of
another well-tested method of computing localization lengths for
disordered systems, the decimation method \cite{LamW80} where one
replaces the full Hamiltonian (\ref{eq-ham}) by an effective
Hamiltonian for the doubly-occupied sites only.  In accordance with
the SP case \cite{MacK83}, $\lambda_2$ is defined by the TIP Green
function, ${\bf G_2}(E)$, as \cite{OppWM96}
\begin{equation}
  {1\over\lambda_2} = - {1\over\vert L-1\vert} \ln\vert\langle
  1\vert {\bf G_2}\vert L\rangle\vert.
\label{eq-gf-lambda2}
\end{equation}
We computed \cite{LeaRS99,LeaRS98,RomLS99} the Green function at $E=0$
for 26 disorder values $W\in[0.5, 9]$, for 24 system sizes $L\in [51,
251]$, and 11 interaction strengths $U= 0, 0.1, \ldots, 1.0$.
For each triplet of parameters $(W,L,U)$ we averaged $1/\lambda_2$
over $1000$ samples.  By FSS of the $\lambda_2(U)$ data we obtained
the infinite-size localization lengths $\xi_2(U)$ shown in
Fig.~\ref{fig-tipdm-xi0}.

We fitted our results by various suggested models. The best
fit was obtained with
\begin{equation}
  \xi_2(U) \propto \xi_2(0)^{\beta(U)}
   \left[ 1 + \frac{c}{\xi_2(0)} \right].
\label{eq-ps}
\end{equation}
\begin{figure}[t]
\begin{minipage}[b]{\halffigwidth}
  \epsfxsize=\fullfigwidth 
  \epsffile{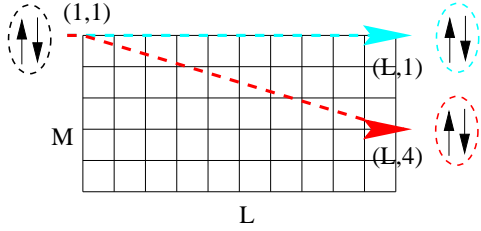}
\end{minipage}\quad\hfill
\begin{minipage}[b]{\halffigwidthb}
\caption{\label{fig-tip2d-diagram}
  Schematic view of two possible TIP decays in the 2D bar
  geometry.}\vspace*{0.5cm}
\end{minipage}
\end{figure}
Such a $U$-dependent exponent for the enhancement $\xi_2(U)/\xi_2(0)$
had been previously predicted \cite{PonS97} with $\beta$ up to 2 for
$U=1$.  However, we find that $\beta < 1.5$ as shown in the inset
of Fig.~\ref{fig-tipdm-xi0}. Thus we do not see the quadratic
behavior (\ref{eq-lambda2}) when using the fit function of Ref.\ 
\cite{PonS97}.  On the other hand, after scaling the data onto a
single scaling curve and fitting
\begin{equation}
  \xi_2(U) - \xi_2(0) \propto \xi_2(0)^\beta
\label{eq-oppen}
\end{equation}
as proposed with $\beta=2$ \cite{OppWM96}, we find \cite{LeaRS99}
indeed $\beta=2$ for not too small disorder, e.g., $W \geq 2.5$ for
$U=1$, but observe a crossover to $\beta=3/2$ for smaller $W$.  

For $U
\gtrsim 1.5$ the enhancement decreases again \cite{WaiWP99,PicWDJ98};
the position of the maximum depends upon $W$. We thus find that the
duality proposed \cite{WaiWP99} for results at $U$ and $\sqrt{24}/U$
--- which would imply a fixed maximum at $U={24}^{1/4}$ --- is
approximately valid. 

We remark that similar results are obtained when
placing TIP in different potentials \cite{LeaRS98} which is of
relevance for a proposed experimental test of the TIP effect
\cite{BriGKT98}.

\section{Localization-delocalization transition for TIP in 2D}
\label{sec-tip-dm2d}

Imry \cite{Imr96,Imr95} argued that two (onsite-)interacting particles
in a 2D random potential show enhancement of $\lambda_2$ at $E=0$ as
\begin{equation}
\lambda_2
 \propto    \lambda_1 \exp\left[
                       \frac{U^{2}\lambda_1^{2}}{t^2}
                      \right]
 \quad\gg\quad
            \lambda_1 \propto \exp \left[\frac{t^2}{W^2}\right],
\label{eq-gf-lambda2-2d}
\end{equation}
with $\lambda_1$ the SP localization length in 2D \cite{MacK83}.
Indeed, Ortu\~{n}o and Cuevas \cite{OrtC99} numerically find a
distinct enhancement in 2D for two particles with a short-ranged
interaction. It is even so strong that an MIT occurs for $U=1$ at
$W_c= 9.3 \pm 0.2$ with critical exponent $\nu = 2.4 \pm 0.5$. Their
result is based on the recursive Green function method previously
employed for the 1D TIP case \cite{OppWM96}. 

We applied the decimation
method to the 2D TIP problem for rectangular bars of size $M\times L$
with $M < L$ at $E=0$ for $36$ disorders $W\in [3.5, 12]$, for system
widths $M= 2, \ldots, 8$ at fixed $L=51$, and $51$ interactions
strengths $U= 0, 0.04, \ldots, 2.0$ with hard wall boundary
conditions.
\begin{figure}[t]
  \epsfxsize=\figwidth \centerline{\epsffile{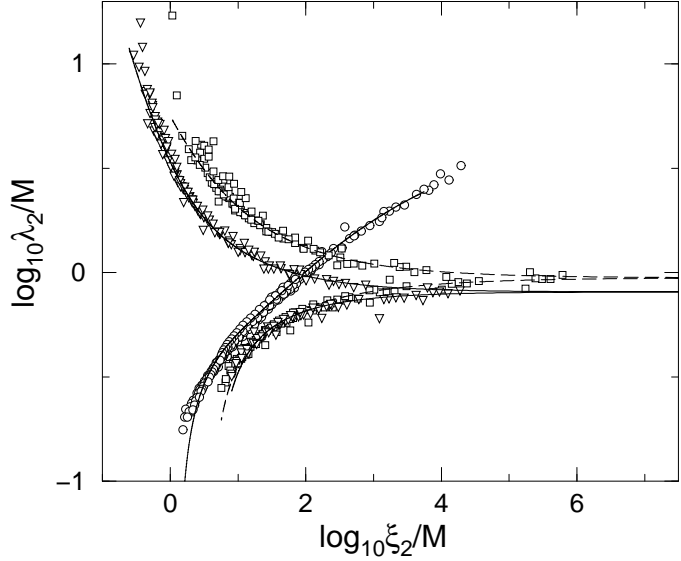}}
 \caption{\label{fig-tip2d-fss}
   FSS scaling curves (lines) and localization lengths $\lambda_2$ for
   TIP in 2D as a function of the scaling parameter $\xi_2$ for $U=0$
   ($\circ$), $1$ ($\Box$), and $2$ ($\nabla$).}
\end{figure}
For the results presented here, we average the decay length of TIP
states over every possible transmission path between coordinates
$(1,1)$ and various heights $y$ on the right hand side $(L,y)$ of the
bar as shown in Fig.~\ref{fig-tip2d-diagram}
\begin{equation}
\frac{1}{\lambda_2}
= -\left\langle \frac{1}{M}
    \sum_{y=1}^{M} 
     \frac{\ln\vert\langle 1,1\vert {\bf G_2}\vert L,y\rangle\vert}%
          {\sqrt{(L-1)^2 + (y-1)^2}}
  \right\rangle_W.
\label{eq-lambda2-2da}
\end{equation}
Here $\langle\cdot\rangle_W$ denotes the disorder average over $1500,
500, 250, 200, 100, 50$ and $20$ configurations for $M=2, \ldots, 8$,
respectively \cite{geometry}. To obtain a more accurate result the
value of $1/\lambda_2$ was actually obtained by fitting the slope
of $\left\langle \ln\vert\langle 1,1\vert {\bf G_2}\vert
  x,y\rangle\vert \right\rangle$ versus $\sqrt{(x-1)^2+(y-1)^2}$
where $x\leq L$.  

In Fig.~\ref{fig-tip2d-fss}, we show typical FSS
results for three values of $U$.
\begin{figure}[t]
  \epsfxsize=\figwidth \centerline{\epsffile{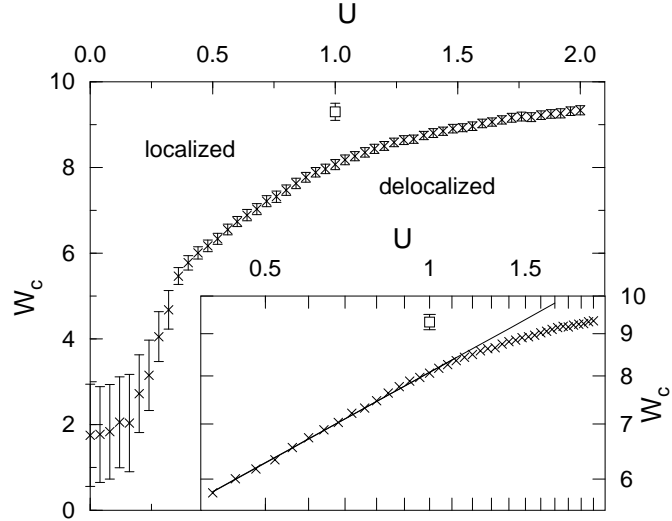}}
 \caption{\label{fig-tip2d-wc}
   Phase diagram of the TIP localization-delocalization transition in
   2D.  The data point ($\Box$) for $U=1$ indicates the result of
   Ref.\ \protect\cite{OrtC99}. The error bars represent the error
   propagation of the Levenberg-Marquardt nonlinear fit. Inset:
   power-law fit to the data for $U\geq 0.4$ with exponent $0.36$. } 
\end{figure}
The quality of the scaling curves is not as good as in the 1D TIP
analysis \cite{LeaRS99}, due to the smaller samples and smaller number of
configurations.  The scaling curves for $U\lesssim 0.2$ can be
described by a single branch corresponding to localized behavior. For
$U\gtrsim 0.5$, the scaling curves have two branches
indicating a transition of TIP states from localized to delocalized
behavior. In the intermediate range $0.2 < U <0.5$, FSS is less
convincing and requires further investigations.  We remark that the
FSS results do not change appreciably within the accuracy of the data
when we drop the data for $M=2$ or $8$.

\begin{figure}[t]
  \epsfxsize=\figwidth \centerline{\epsffile{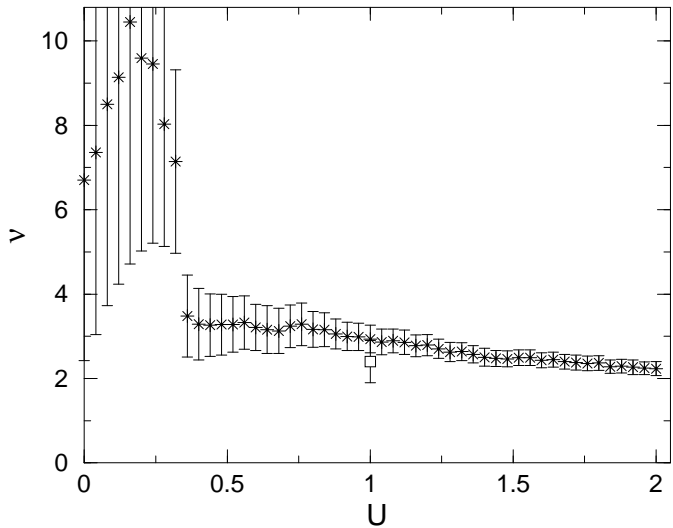}}
 \caption{Critical exponent $\nu$ according to Eq.\ (\ref{eq-fit}).
   The data point ($\Box$) for $U=1$ represents the result of Ref.\ 
   \protect\cite{OrtC99}. The error bars are obtained as in
   Fig.~\protect\ref{fig-tip2d-wc}.}
 \label{fig-tip2d-nu}
\end{figure}

In order to reliably extract the critical disorders $W_c$ and the
correlation length exponents $\nu$ from the FSS data, we employed the
FSS techniques of Refs.\ \cite{MacK83} and \cite{SleO99a}. In the
lowest order approximation, the fit function is given as \cite{MacK83}
\begin{equation}
\frac{\lambda_2}{M}
= \Lambda_{c} + a M^{1/\nu}\left(\frac{W}{W_c}-1\right),
\label{eq-fit}
\end{equation}
with fit parameters $\Lambda_{c}$, $a$, $W_c$ and $\nu$. We used
a fit function containing up to the third power in ${W}/{W_c}$
\cite{SleO99a}. In this way including deviations from the linear
behavior in $W$ is important due to limited quality of the data. In
Figs.\ \ref{fig-tip2d-wc} and \ref{fig-tip2d-nu}, we show the
resulting dependence of $W_c$ and $\nu$ on $U$.
Note how the FSS gives unreliable results for $U < 0.5$, resulting in
the incorrect prediction of $W_c(0) > 0$. As demonstrated in the inset
of Fig.~\ref{fig-tip2d-wc} the $W_c$ data can be approximated for $0.4
\leq U \leq 1$ by a power-law fit $W_c \propto U^{0.36\pm 0.03}$.  The
critical exponent $\nu$ shows a slight decrease for $U \geq 0.5$
and is in reasonable agreement with the value of $2.4 \pm 0.5$
obtained in Ref.\ \cite{OrtC99} for $U=1$.  However, the
(non-universal) value of $W_c=8.1\pm 0.1$ at $U=1$ is somewhat
different Ref.\ \cite{OrtC99}. We attribute this to our different
definition (\ref{eq-lambda2-2da}) of $\lambda_2$ \cite{geometry}.

\section{Conclusions}
\label{sec-tip-concl}

In conclusion, we observe an enhancement of the TIP localization
lengths due to onsite interaction both in 1D and 2D random potentials.
This enhancement persists, unlike for TMM
\cite{RomS97a,RomS98,HalMK98}, in the limit of large system size and
after constructing infinite-sample-size estimates from the FSS curves.
In the 2D case, it leads to numerical evidence for a
localization-delocalization transition of the TIP states. We
emphasize that this is not a metal-insulator transition in the
standard sense since only the TIP states show the delocalization
transition. The majority of non-paired states remains localized.

\vspace*{0.25cm} \baselineskip=10pt{\small \noindent RAR gratefully
  acknowledges support by the DFG via SFB-393. ML gratefully acknowledges
  the support of the EU-TMR programme under grant no.\ ERBFMBICT972832.}
%
%
%
%
%
%
%
%
%
%
%
%


\begin{thebibliography}{10}
\bibitem{geometry} The geometric factor in the denominator in
  Eq.~(\ref{eq-lambda2-2da}) has not been taken into account in Ref.\ 
  \cite{OrtC99}
\end{thebibliography}

\end{document}